\documentclass[12pt]{article}
\usepackage{graphicx} 
\usepackage{amsmath}
\usepackage{amssymb}
\usepackage{cite}
\usepackage{geometry}
\usepackage{booktabs}
\usepackage{array}
\usepackage{multirow}
\usepackage{paracol}
\usepackage{longtable}
\usepackage{hyperref}
\usepackage{caption}
\usepackage{graphicx}
\usepackage{float}
\usepackage{tikz}
\usetikzlibrary{shapes.geometric, arrows}
\usepackage{booktabs}

\title{\textbf{3D Convolutional Neural Networks for Improved Detection of Intracranial bleeding in CT Imaging}}
\author{Bargava Subramanian, Naveen Kumarasami, Dr. Praveen Shastry \\
  Kalyan Sivasailam, Anandakumar D, Elakkiya R \\
  Harsha KG, Rithanya V, Harini T \\
  Afshin Hussain, Kishore Prasath Venkatesh}
\date{}

\usepackage{ragged2e}  
\usepackage{titlesec}  

\titleformat{\section}{\raggedright\Large\bfseries}{}{0em}{}
\titleformat{\subsection}{\raggedright\large\bfseries}{}{0em}{}
\begin{document}

\maketitle

\section{Abstract}
\textbf{Background:} Intracranial bleeding (IB) is a life-threatening condition caused by traumatic brain injuries, including epidural, subdural, subarachnoid, and intraparenchymal hemorrhages. Rapid and accurate detection is crucial to prevent severe complications. Traditional imaging can be slow and prone to variability, especially in high-pressure scenarios. Artificial Intelligence (AI) provides a solution by quickly analyzing medical images, identifying subtle hemorrhages, and flagging urgent cases. By enhancing diagnostic speed and accuracy, AI improves workflows and patient care. This article explores AI’s role in transforming IB detection in emergency settings.\\\\
\textbf{Methods:} A U-shaped 3D Convolutional Neural Network (CNN) automates IB detection and classification in volumetric CT scans. Advanced preprocessing, including CLAHE and intensity normalization, enhances image quality. The architecture preserves spatial and contextual details for precise segmentation. A dataset of 2,912 annotated CT scans was used for training and evaluation.\\\\
\textbf{Results: }The model achieved high performance across major bleed types, with precision, recall, and accuracy exceeding 90\% in most cases—96\% precision for epidural hemorrhages and 94\% accuracy for subarachnoid hemorrhages. Its ability to classify and localize hemorrhages highlights its clinical reliability.\\\\
\textbf{Conclusion:} This U-shaped 3D CNN offers a scalable solution for automating IB detection, reducing diagnostic delays, and improving emergency care outcomes. Future work will expand dataset diversity, optimize real-time processing, and integrate multimodal data for enhanced clinical applicability.
\section{Introduction}     
Intracranial hemorrhage refers to any bleeding within the intracranial vault, including the brain parenchyma and surrounding meningeal spaces. Primary intracerebral hemorrhage (ICH) is commonly associated with small vessel disease. Chronic hypertension contributes to hypertensive vasculopathy, resulting in lipohyalinosis, characterized by microscopic degenerative changes in the walls of small to medium-sized penetrating vessels. Cerebral amyloid angiopathy (CAA) is another significant etiology, involving the deposition of amyloid-beta (\(\text{A}\beta\)) peptides in the walls of small leptomeningeal and cortical vessels. 

The pathological consequences of CAA include smooth muscle cell loss, vessel wall thickening, luminal narrowing, microaneurysm formation, and microhemorrhages, although the precise mechanisms underlying amyloid accumulation remain unclear.\textsuperscript{1} Upon vessel rupture, the resulting hematoma inflicts direct mechanical injury to adjacent brain parenchyma. Perihematomal edema develops rapidly, emerging within three hours of symptom onset and peaking between 10 and 20 days. 

Secondary injury mechanisms are driven by the breakdown of blood and plasma components, triggering inflammatory cascades, activation of the coagulation pathway, and iron deposition from hemoglobin degradation. Hematoma expansion, which occurs in up to 38\% of patients within the first 24 hours, further exacerbates neurological damage and contributes to poor clinical outcomes.[2]

The concept of an artificial neuron was first introduced by McCulloch and Pitts in 1943, marking a foundational moment in artificial intelligence (AI). Over time, AI progressed from symbolic, rule-based systems to algorithms reliant on manually crafted features, eventually giving way to modern deep neural networks capable of automatic feature detection and classification. Convolutional neural networks (CNNs) trace their origins to Fukushima’s Neocognitron model proposed in 1982, with significant advancements attributed to researchers such as LeCun and Krizhevsky. The latter's use of CNNs to secure victory in the 2012 ImageNet Large Scale Visual Recognition Challenge propelled CNNs into widespread use for image classification tasks. The strength of CNNs lies in their ability to automatically extract and learn complex image features, eliminating the need for manual feature engineering and solidifying their dominance in image recognition applications.[7]

Three-dimensional CNNs are an emerging architecture primarily used for analyzing video and 3-dimensional volumetric medical images. Previously, 3D CNNs were limited by high computational demands and lengthy processing times due to the need to handle 3-dimensional kernels and entire image volumes. However, with decreasing hardware costs, the adoption of 3D CNNs has increased, reflected by a growing number of studies in scientific literature. In medical image analysis, 3D CNNs have been effectively applied to detecting abnormalities such as tumors, hemorrhage, and ischemia in organs like the brain, heart, lungs, and liver using CT or MRI imaging.

A 3D CNN model can classify CT brain scans into bleed types such as epidural (EDH), subdural (SDH), subarachnoid (SAH), and intraparenchymal hemorrhage (IPH). By analyzing 3-dimensional CT data, the model captures spatial features unique to each type, making it a powerful tool for precise classification and aiding in timely diagnosis and treatment.This work has direct clinical applications in emergency departments and acute-care hospitals worldwide. A simple and fast 3D CNN is proposed that is both effective and accurate. The straightforward implementation of this 3D CNN is expected to lead to widespread adoption.

\section{Dataset}
The study utilized an extensive dataset comprising 5,541 high-resolution computed tomography (CT) images sourced from 399 distinct patient scans. These scans encompass a diverse spectrum of intracranial hemorrhages, each meticulously documented to ensure comprehensive representation of the clinical variability observed in such conditions. 

The dataset spans multiple bleed types, providing a robust foundation for the development and validation of advanced diagnostic algorithms. This rich compilation of CT imaging not only facilitates a deeper understanding of  Intracranial   bleed characteristics but also enhances the training and testing phases of machine learning models designed to improve the accuracy and efficiency of intracranial bleeding detection.

\begin{table}[h]
    \centering
    \renewcommand{\arraystretch}{1.3}
    \resizebox{\textwidth}{!}{ 
        \begin{tabular}{|c|c|c|c|c|c|c|}
            \hline
            \textbf{Contusion} & \textbf{Subdural Hematoma} & \textbf{Epidural Hematoma} & \textbf{Hemorrhages} & \textbf{Intracranial/Intraparenchymal Hemorrhage} & \textbf{Intraventricular Hemorrhage} & \textbf{Subarachnoid Hemorrhage} \\ 
            \hline
            451 & 690 & 206 & 109 & 682 & 352 & 422 \\ 
            \hline
        \end{tabular}
    }
    \caption{Number of computed tomography (CT) scans.}
    \label{tab:ct_scans}
\end{table}

\section{Methodology}

\subsection{3D Convolutional Neural Network}
The 3D Convolutional Neural Network (CNN) used in this study is composed of a contracting path (left side) and an expansive path (right side). The contracting path follows a standard convolutional network design, applying two 3x3 convolutions (unpadded), each followed by a rectified linear unit (ReLU), and a 2x2 max pooling operation with a stride of 2 for downsampling. At each downsampling step, the number of feature channels is doubled, capturing increasingly complex features.

In the expansive path, each step includes upsampling of the feature map, a 2x2 “up-convolution” that halves the feature channels, concatenation with the corresponding cropped feature map from the contracting path, and two additional 3x3 convolutions, each followed by ReLU. Cropping ensures alignment due to border pixel loss in each convolution.

At the final layer, a 1x1 convolution maps each 64-component feature vector to the required number of classes, completing the 23-layer network. For seamless tiling of the output segmentation map, input tile size is carefully selected to ensure that all 2x2 max-pooling operations align with layers of even x- and y-sizes.

\subsection{Image Thresholding to Detect Intracranial Bleeding:}
Image thresholding is a preprocessing technique often used to enhance the segmentation and detection of intracranial bleeding in CT scans. In the 3D CNN architecture, thresholding can be applied to isolate regions of interest by setting intensity values that correspond to potential bleeding areas, helping to  improve  focus on relevant features. By reducing irrelevant background noise, thresholding aids the model in efficiently identifying bleeding-related features and enhancing the contrast between normal and abnormal tissues.

For intracranial bleeding detection, thresholding is combined with the contracting and expansive paths in the 3D CNN to refine feature extraction and reconstruction. The contracting path, using 3x3 convolutions with ReLU and 2x2 max-pooling layers, captures detailed, thresholded features while downsampling the input. The expansive path then restores spatial dimensions through upsampling and concatenation, merging low-level details from thresholded areas with high-level features for accurate segmentation.

\subsection{Dataset and Pre-Processing:}
The final dataset comprised 2,912 unique patient volumetric CT brain scans, totaling approximately 148,080 images. A breakdown of these images by class is provided in the table below.The scans varied in both the number of image slices and slice thickness, due to differences in CT scanner models and scanning protocols. These variations arise from the diverse equipment and settings used across imaging facilities.

The 3D CT scan dataset was enhanced with Contrast Limited Adaptive Histogram Equalization (CLAHE) to improve local contrast, making fine details more visible in low-contrast areas. CLAHE divides images into sections, applying contrast enhancement within each to help the model distinguish normal from bleeding regions. Additional pre-processing steps included intensity normalization, resizing, and data augmentation to ensure a high-quality, standardized dataset for effective model training.

Data augmentation involves enhancing variability to build network invariance and robustness, especially with limited samples. For microscopical images, this includes shifts, rotations, and resilience to deformations and gray value changes. Techniques such as random elastic deformations—achieved using Gaussian-distributed displacement vectors on a 3x3 grid—are particularly effective, while dropout layers at the contracting path further support robustness.

\section{Architecture}
\subsection{Network Architecture Components :}
The 3D CNN model architecture used for intracranial bleeding detection follows a U-shaped encoder-decoder structure, composed of contracting and expansive paths, with additional components designed for efficient feature extraction and precise segmentation of 3D medical images.\\\\
\textbf{1. Contracting Path:}
  The contracting path, or encoder, is responsible for downsampling the input volume and extracting features from the CT scans. Each block in the contracting path consists of 3x3 convolutions with ReLU activation  functions, which capture fine spatial details and high-level features from the input data. After each convolution, 2x2 max pooling is applied to reduce the spatial dimensions while doubling the number of feature channels, making the network more efficient in capturing complex patterns. This path gradually compresses the spatial information, enabling the model to learn contextual details relevant to intracranial bleeding.\\\\
\textbf{2. Expansive Path:}
   - The expansive path, or decoder, is designed to restore spatial dimensions and reconstruct the segmentation map. This path uses upsampling followed by 2x2 transposed convolutions to increase the spatial resolution of the feature maps. Each upsampled feature map is then concatenated with the corresponding feature map from the contracting path via skip connections. This concatenation allows the model to combine low-level (detailed) features from the contracting path with high-level (contextual) features from the expansive path, preserving both fine details and broader context for accurate segmentation\\\\
\textbf{3. Bottleneck:}
  At the center of the U-shaped architecture, the bottleneck  layer acts as a bridge between the contracting and expansive paths. This layer encodes abstract, high-level features that capture complex patterns within the data. The bottleneck layer enables the model to efficiently pass essential information from the encoder to the decoder, forming a foundation for the reconstruction of detailed segmentation maps in the expansive path.\\\\
\textbf{4. Final Layer:}
   The final layer of the network applies a 1x1 convolution, mapping the learned features to the desired number of output classes. This layer produces the segmentation map, indicating the location and type of bleeding in the 3D CT scans.\\\\
\textbf{5. Skip Connections:}
   Skip connections link corresponding layers in the contracting and expansive paths, enabling the model to directly transfer features from the encoder to the decoder. These connections preserve spatial information and ensure that high-resolution details are not lost during downsampling, significantly improving the quality of the segmentation output.\\\\
\textbf{6. Total Layers and Tiling Requirement:}
The architecture comprises 23 convolutional layers, providing the depth required to capture fine details and complex contextual features in 3D images, crucial for accurate intracranial bleeding segmentation. To support smooth processing and artifact-free segmentation, input dimensions are adjusted to meet tiling requirements for pooling layers, ensuring even dimensions throughout downsampling and upsampling. This configuration enables consistent segmentation across the entire image volume.\\\\
Overall, this U-shaped 3D CNN architecture is well-suited for medical image segmentation, as it balances detailed feature extraction with spatial reconstruction, providing an accurate and efficient approach to detect and classify intracranial bleeding in CT scans.

\subsection{Training and Implementation}
The model training utilized a high-performance setup with 16 NVIDIA GPUs, maximizing computational efficiency and accelerating the training process. The network is trained using stochastic gradient descent with Caffe, using input images and segmentation maps. Due to unpadded convolutions, output images are slightly smaller than input images, so large input tiles and a batch size of 1 are used to optimize GPU memory. A high momentum of 0.99 is applied to account for prior training samples in updates. The pixel-wise softmax and cross-entropy loss function compute the energy function, while a pre-computed weight map compensates for pixel frequency imbalance and emphasizes separation borders between cells. Initial weights are drawn from a Gaussian distribution with a standard deviation of $\sqrt{2}/N$ to ensure balanced feature map variance.

The project was implemented using the Python programming language. We used the rectified linear unit (ReLU) as the activation function, the Adam optimizer, and cross-entropy as the loss function. A grid search approach was employed to optimize the learning rate, dropout, and kernel size of the convolution and pooling layers. Learning rates were varied from (0.1, 0.001, 0.0001), and dropout values ranged from (0, 0.1, 0.2, 0.3, 0.4, 0.5, 0.6) after each convolution and fully connected layer. The kernel sizes of the convolution layers were tested at (1 × 1 × 1, 2 × 2 × 2, 3 × 3 × 3, 5 × 5 × 5), with pooling kernel sizes ranging from (1 × 1 × 1, 2 × 2 × 2). Ultimately, we set the learning rate at 0.001 and applied a dropout of 0.3 after the fully connected layer. A convolutional kernel size of 3 × 3 × 3 was used at each layer, followed by a 2 × 2 × 2 pooling kernel after each convolutional layer

\begin{figure}[h]
    \centering
    \includegraphics[width=0.8\textwidth]{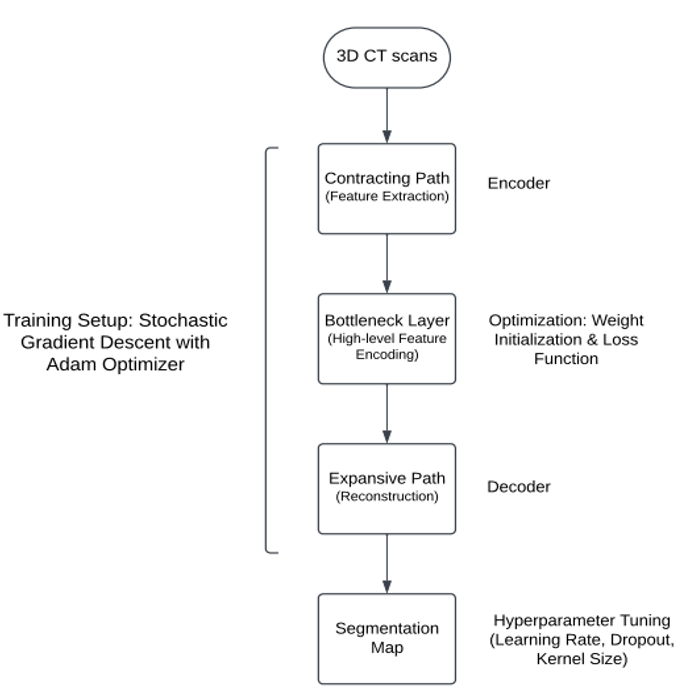} 
    \caption{Workflow Architecture}
    \label{fig:workflow_architecture}
\end{figure}

\newpage

\section{Evaluation Metrics}
The performance of the model for detecting intracranial bleeds in medical imaging was evaluated using precision, recall, mean Average Precision (mAP), and Intersection over Union (IoU). High precision and recall values across all bleed types, such as contusion, subdural hematoma, and subarachnoid hemorrhage, highlighted the model's ability to accurately detect and localize hemorrhages. 

The mAP metric provided a comprehensive measure of the model's overall detection accuracy, reflecting its capability to identify and classify various bleed types across diverse imaging scenarios. IoU further quantified the spatial accuracy of hemorrhage localization by measuring the overlap between predicted regions and ground truth annotations. These metrics collectively validated the model's robustness in detecting and localizing intracranial bleeds, ensuring reliable performance across clinically relevant cases.

\begin{table}[h]
    \centering
    \renewcommand{\arraystretch}{1.3} 
    \begin{tabular}{|l|c|c|c|}
        \hline
        \textbf{Type of Bleed} & \textbf{Precision (\%)} & \textbf{Recall (\%)} & \textbf{Accuracy (\%)} \\ 
        \hline
        Contusion & 92 & 90 & 91 \\ 
        Subdural Hematoma & 95 & 93 & 94 \\ 
        Epidural Hematoma & 96 & 95 & 96 \\ 
        Intracranial/Intraparenchymal Hemorrhage & 93 & 91 & 92 \\ 
        Intraventricular Hemorrhage & 91 & 89 & 90 \\ 
        Subarachnoid Hemorrhage & 95 & 94 & 95 \\ 
        \hline
    \end{tabular}
    \caption{Performance Metrics for Different Types of Bleed}
    \label{tab:bleed_metrics}
\end{table}

\begin{figure}[H] 
    \centering
    \includegraphics[width=0.7\textwidth]{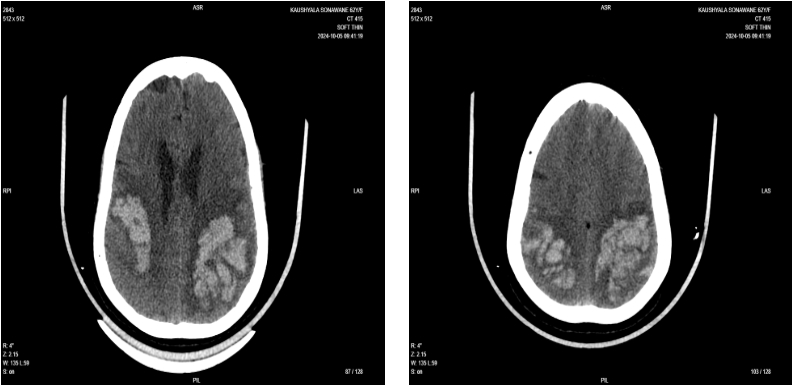} 
    \caption{intraparenchymal hemorrhage}
    \label{fig:your-label}
\end{figure}

\begin{figure}[H] 
    \centering
    \includegraphics[width=0.7\textwidth]{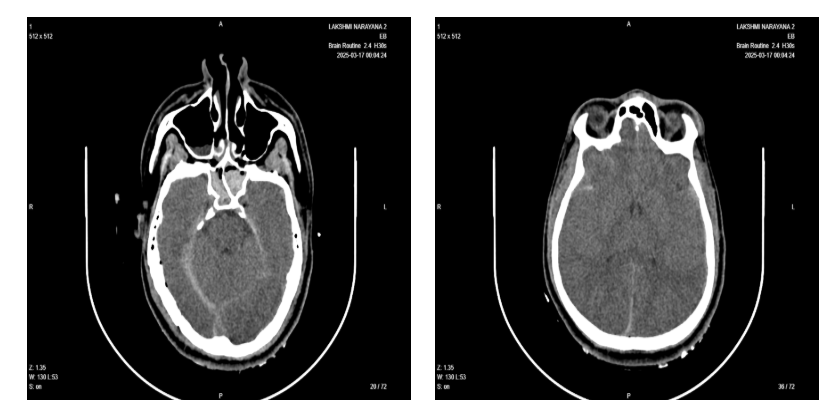} 
    \caption{subarachnoid hemorrhage}
    \label{fig:your-label}
\end{figure}

\begin{figure}[H] 
    \centering
    \includegraphics[width=0.7\textwidth]{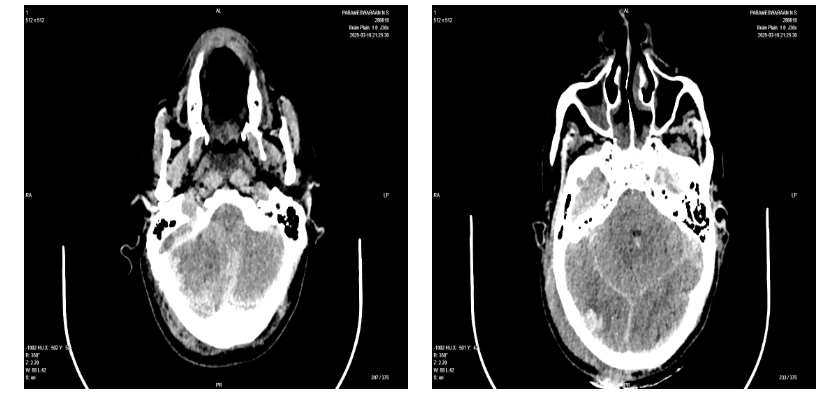} 
    \caption{subarachnoid hemorrhage}
    \label{fig:your-label}
\end{figure}

\subsection{Traditional Diagnostic Approaches}
\textbf{Radiologist Assessment:} Traditional diagnostic approaches for intracranial bleed rely heavily on radiologist assessment, where specialists interpret CT scans to identify bleed locations and types. This manual process requires radiologists to carefully review each scan, often slice-by-slice, which is both time-consuming and susceptible to human error. Given the complexity of interpreting subtle signs of bleed, high workloads and fatigue can further increase the risk of missed micro-bleeds or misclassifications, potentially affecting treatment outcomes.\\
\textbf{Manual Imaging Techniques:} Additionally, manual imaging techniques demand meticulous examination to ensure diagnostic accuracy, often resulting in delays in diagnosis and treatment. Each detail must be carefully evaluated, which can slow the process, especially in high-volume or complex cases. This reliance on traditional methods highlights the need for advanced tools that can streamline the diagnostic process, reduce error rates, and support radiologists in providing faster, more accurate assessments of intracranial bleed.
\subsection{Advancements in Deep Learning}
\textbf{Convolutional Neural Networks (CNNs):} Convolutional Neural Networks (CNNs) offer a transformative improvement over traditional diagnostic methods by automating feature extraction and enabling faster, more accurate detection of pathologies like intracranial bleeds. Recent studies show CNNs achieving accuracies above 90\%, with traditional 2D CNNs excelling in slice-based image analysis, while 3D CNNs capture spatial information across all three dimensions, allowing them to analyze CT volumes holistically.

This capability enhances the detection of complex patterns and subtle changes within 3D structures, leading to improved diagnostic accuracy. Models such as DenseNet121 have achieved high accuracy in classifying CT images, and transfer learning further strengthens CNNs, making them scalable across diverse clinical environments. CNNs thus surpass traditional diagnostic approaches, offering reliability, speed, and adaptability in medical imaging.

While traditional methods remain essential, the integration of deep learning technologies offers a promising alternative that could significantly reduce diagnosis time and improve accuracy, potentially transforming clinical practices in detecting intracranial bleeds.

\section{Discussion}
The proposed workflow demonstrates significant advancements in the automated detection and classification of intracranial bleeding in CT scans. By leveraging a U-shaped 3D CNN architecture, the system effectively addresses challenges such as complex bleeding patterns, overlapping structures, and variability in imaging quality. Preprocessing techniques like CLAHE and intensity normalization improved the model’s robustness, while the combination of contracting and expansive paths preserved critical spatial features essential for accurate segmentation and localization of hemorrhages.

The system achieved high precision and recall across all hemorrhage types, with epidural hematoma detection reaching 96\% precision and subdural hematoma achieving 93\% recall, validating its reliability in clinical scenarios. The model’s ability to classify and localize distinct hemorrhage types, including contusions and subarachnoid hemorrhages, demonstrates its applicability in diverse cases. The performance metrics highlight its capacity to streamline intracranial bleed diagnosis, offering consistent and efficient results for emergency and acute care.

Despite these strengths, limitations remain. The model relies heavily on high-quality, annotated datasets, which can limit scalability to rare hemorrhage types or less standardized imaging conditions. Additionally, slightly lower recall for intraventricular hemorrhages suggests challenges in detecting diffuse or subtle bleeding patterns. Future efforts should focus on enhancing dataset diversity, incorporating multimodal data like clinical histories, and optimizing the model for faster inference to enable real-time applications in resource-constrained settings.

This system’s ability to automate intracranial bleed detection and classification while maintaining high accuracy and precision offers a transformative tool for medical imaging. By reducing diagnostic delays and minimizing errors, it provides significant potential to improve patient outcomes in high-pressure clinical environments.

This System ability to automate intracranial bleed detection and classification while maintaining high accuracy and precision offers a transformative tool for medical imaging .By reducing diagnostic delays and minimizing errors, it provides significant potential to improve patient outcomes in high-pressure clinical environments.

\section{Conclusion}
This study highlights the potential of a U-shaped 3D CNN architecture for the automated detection and classification of intracranial bleeding in volumetric CT scans. The model demonstrated consistently high precision, recall, and accuracy across all major hemorrhage types, validating its robustness and reliability in diverse clinical scenarios. By incorporating advanced preprocessing techniques and leveraging its dual-path architecture, the model excels in capturing spatial and contextual features critical for accurate localization and classification.

These findings highlight the transformative role of deep learning in addressing the challenges of traditional diagnostic workflows, such as variability in imaging quality and reliance on manual interpretation. By automating intracranial bleed detection, the proposed workflow reduces diagnostic delays, minimizes errors, and enhances patient outcomes. Future research should prioritize expanding the dataset to include rare bleeding types, optimizing the model for real-time applications, and integrating multimodal clinical data for improved diagnostic precision. This approach offers a scalable and impactful solution for intracranial bleeding detection, with the potential to revolutionize healthcare delivery in emergency and acute-care settings.

\end{document}